\documentclass[fp,twocolumn]{jpsj3}

\usepackage{color}
\usepackage{braket} 
\usepackage{txfonts}
\usepackage{ulem}
\usepackage{multirow}
\usepackage{bm}

\title{Mean-field Study of Antiferromagnetic and Antiferroquadrupolar Orderings \\ 
in Tetragonal CeCoSi}

\author{Megumi Yatsushiro$^{1,2}$, Satoru Hayami$^{1,2}$}
\inst{$^1$Department of Physics, Hokkaido University, Sapporo 060-0810, Japan \\
 $^2$ Department of Applied Physics, The University of Tokyo, Tokyo 113-8656, Japan} 

\abst{
We investigate the stability of the multipolar orderings in $f$-electron material CeCoSi based on a self-consistent mean-field calculation for the effective localized model.
This material has two ordered phases in the temperature-pressure phase diagram: the antiferromagnetic phase and the nonmagnetic phase, the latter of which has been suggested to be an antiferroquadrupolar phase. 
Meanwhile, the origin of the antiferroquadrupolar phase has been unclear, since a quadrupole degree of freedom is present only between the ground-state level and highly separated excited-state level under a tetragonal crystalline electric field (CEF), whose energy scale is much larger than the transition temperature.
To understand the sequence of the phase transition from the paramagnetic phase, antiferroquadrupolar phase, and antiferromagnetic phase when decreasing the temperature, we examine the important interaction parameters in the effective localized model.
We clarify that the $3z^2-r^2$-type of the antiferroquadrupolar interactions can renormalize the CEF level splitting, which 
might assist a quadrupolar ordering even in a tetragonal system without orbital degeneracy.
Moreover, the stability of the antiferroquadrupolar and antiferromagnetic states in a magnetic field and the behavior of the magnetic/quadrupolar susceptibility are also presented for the information to identify the unknown order parameter in the nonmagnetic ordered phase.
}

\begin{document}
\maketitle

\section{Introduction \label{sec:intro}}

An antiferroquadrupolar (AFQ) ordering is one of the typical multipolar ordered states in $4f$, $5d$, and $5f$ electron systems~\cite{kusunose2008description, kuramoto2009multipole, Santini_RevModPhys.81.807, suzuki2018first}. 
Such AFQ orderings have been often found in high-symmetry cubic systems with the orbital degeneracy in the crystalline electric field (CEF) ground state so as to include electric quadrupole degrees of freedom, e.g., CeB$_6$~\cite{takigawa1983nmr, luthi1984elastic, EFFANTIN1985145, ERKELENS198761, nakamura1994quadrupole, sakai1997new, shiina1997magnetic, shiina1998interplay, doi:10.1143/JPSJ.70.1857, cameron2016multipolar}, PrPb$_3$~\cite{MORIN1982257, doi:10.1143/JPSJ.70.248, doi:10.1143/JPSJ.73.2377, PhysRevLett.94.197201,doi:10.1143/JPSJS.77SA.199, doi:10.1143/JPSJ.79.093708, PhysRevB.95.054425}, Pr$T_2X_{20}$ ($T={\rm Ir, Rh}, X={\rm Zn}; T={\rm V}, X={\rm Al}$)~\cite{PhysRevLett.106.177001, doi:10.1143/JPSJ.80.093601, doi:10.1143/JPSJ.80.063701, PhysRevB.86.184426, PhysRevB.87.205106, PhysRevLett.113.267001,doi:10.7566/JPSJ.85.082002,PhysRevB.95.155106}, Ba$_2$MgReO$_6$~\cite{PhysRevResearch.2.022063}, and UPd$_3$~\cite{Andres1978AnomalousBI,MCEWEN1993670,walker1994triple, mcewen1998quadrupolar,PhysRevB.60.R8430, PhysRevLett.87.057201, doi:10.1143/JPSJ.70.1731,PhysRevLett.97.137203,Sikora_2006, Walker_2008}.
On the other hand, the AFQ orderings have been identified in the systems even without the orbital degeneracy in the CEF ground state, e.g., 
cubic CeTe~\cite{kawarasaki2011pressure, doi:10.7566/JPSJ.84.044708, doi:10.7566/JPSCP.3.011035} and tetragonal $R$B$_{2}$C$_{2}$($R={\rm Dy}, {\rm Ho}$)~\cite{doi:10.1143/JPSJ.68.2057, Tanaka_1999, PhysRevLett.84.2706, doi:10.1143/JPSJS.71S.94,  PhysRevB.65.094420, NEMOTO2003641,PhysRevB.69.024417, PhysRevLett.94.036408,doi:10.1143/JPSJ.74.1500, Mulders_2006, 
doi:10.1143/JPSJ.68.2526, doi:10.1143/JPSJ.74.1666, PhysRevB.71.104416, HILLIER2007757, doi:10.1143/JPSJ.81.034715}.

\begin{figure}[t!]
\centering
\includegraphics[width=70mm]{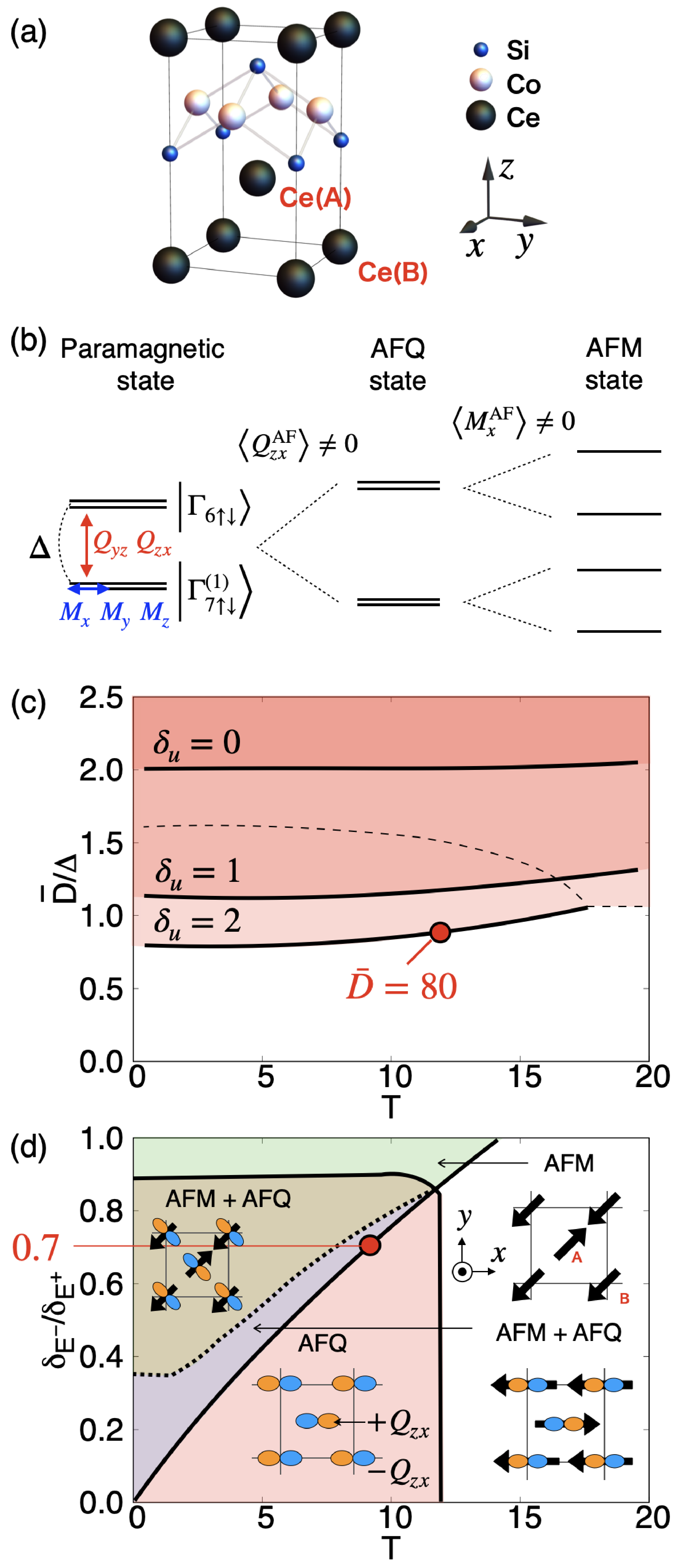}
\caption{(Color online)
(a) Crystal structure of CeCoSi in a unit cell.
(b) Schematic picture of the low-energy four levels in the paramagnetic state (left), AFQ state (middle), and AFM state (right).
(c) The phase diagram while changing the temperature ($T$) and interaction $\bar{D}$ in the absence of the AFM interaction. 
For $\delta_u=0$, $1$, and $2$, the AFQ ordering is stabilized in the colored region above each bold line.
The thin dashed line describes the transition from the $Q_{zx}$($Q_{yz}$)-type AFQ order to the $Q_u$-type one when increasing $\bar{D}$ for $\delta_u=2$.
Other nonzero parameters are $(\delta_{{\rm E}^+}, \delta_\varv)=(1, 0.8)$.
(d) The phase diagram when $T$ and the ratio $\delta_{{\rm E}^-}/\delta_{{\rm E}^+}$ change, where the antiferroic ordering in each phase is schematically presented. The black arrow and colored object stand for the in-plane magnetic moment and electric quadrupole consisting of $Q_{yz}$ and $Q_{zx}$, respectively.
The solid (dotted) line represents the second(first)-order phase transition.
 \label{fig:fig1}}
\end{figure}

CeCoSi is one of the potential candidates with AFQ orderings in the tetragonal crystal structure, where the CEF levels consist of only the Kramers pairs~\cite{chevalier2006antiferromagnetic, lengyel2013temperature, tanida2018substitution, tanida2019successive, kawamura2020structural, nikitin2020gradual, tanida2020magnetic, chandra2020phenomenological, jpsj_yatsushiro2020odd, proc_yatsushiro2020antisymmetric, yatsushiro2020nqr, manago2021unusual,matsumura2022structural, hidaka2022magnetic, kawamura2022structural}.
This material shows the signature of two ordered phases including a higher-rank multipolar ordered phase, which is denoted as II and III phases in addition to a high-temperature paramagnetic phase (I phase); the low-temperature III phase was indicated to be characterized by an in-plane antiferromagnetic (AFM) spin configuration with ordering vector ${\bm q}={\bm 0}$ and the II phase was suggested to be an AFQ phase~\cite{lengyel2013temperature, tanida2018substitution, tanida2019successive, nikitin2020gradual, manago2021unusual}, although the order parameter of the II phase has not been identified yet.
Since the ${\bm q}={\bm 0}$ order described by the staggered-type alignment of the order parameters on two Ce ions [Ce (A) and Ce (B) in Fig.~\ref{fig:fig1}(a)] breaks the global inversion symmetry, parity-violating phenomena like multiferroic and nonlinear responses are expected from the symmetry viewpoint in both AFM and AFQ orderings~\cite{jpsj_yatsushiro2020odd,proc_yatsushiro2020antisymmetric,yatsushiro2020nqr, hayami2018classification,watanabe2018group,PhysRevB.104.054412,doi:10.7566/JPSJ.91.014701}.
Thus, it is highly desired to identify the microscopic order parameters in the II and III phases. 
From this aspect, the authors have recently proposed a way to detect the staggered antiferroic order parameters by the NQR/NMR measurement~\cite{yatsushiro2020nqr}. 
Another intriguing issue is why the II phase is stabilized as the multipolar ordered state in spite of the unprecedentedly large CEF splitting, which has been estimated at around $100$~K~\cite{chevalier2006antiferromagnetic, lengyel2013temperature, tanida2018substitution, tanida2019successive, nikitin2020gradual, tanida2020magnetic}.

In the present study, we investigate the stability of the multipolar orderings of CeCoSi by focusing on the competition between the AFQ and AFM phases at finite temperatures.
Based on a mean-field calculation for an effective localized model, we discuss the magnitude of the multipolar interaction to induce the AFQ phase transition under the large CEF level splitting.
We find that one of the multipole-multipole interactions assists in stabilizing interorbital quadrupolar orderings by renormalizing the CEF level splitting.
Moreover, we examine the magnetic field-temperature phase diagram by aiming at clarifying the key model parameters to reproduce the tendency of the AFQ transition temperature against a magnetic field.
We also present the behaviors of the magnetic and quadrupolar susceptibilities in the ordered phases.
From the analyses in terms of the finite-temperature phase boundaries, we list possible combinations of the antiferroic multipole order parameters in the II and III phases, which will be a reference to identify the order parameters.

This paper is organized as follows. We introduce the effective localized model in Sect.~\ref{sec:model}. 
By using the self-consistent mean-field calculation on the effective model, we examine the stabilities of the AFQ and AFM phases in a zero or finite magnetic field in Sect.~\ref{sec:stability}.
In Sect.~\ref{sec:susceptibility}, we discuss the behaviors of the magnetic and quadrupolar susceptibilities.
After we briefly present the possible order parameters in the AFQ and AFM phases in Sect.~\ref{sec:order_parameter}, we summarize this paper in Sect.~\ref{sec:summary}.
In Appendix~\ref{sec:CEF}, we show the CEF Hamiltonian and level splittings.
We supplementally show the relevance between the interaction and the transition temperature under the highly separated CEF level splitting in Appendix~\ref{sec:D_T0}.

\section{Effective Localized Model \label{sec:model}}
We introduce the effective model for the localized $4f$ electron with $f^1$ configuration in the Ce$^{3+}$ ion under the tetragonal CEF.
In the present calculation, we consider the CEF ground-state $\Gamma_7$ and first-excited $\Gamma_6$ levels [Fig.~\ref{fig:fig1}(b)] by using the CEF parameters proposed by the experiment~\cite{mitsumoto2019} (see also Appendix~\ref{sec:CEF}), although the result for the CEF ground-state $\Gamma_7$ and first-excited $\Gamma_7$ levels is also briefly presented in Table~\ref{table:table1} in Sect.~\ref{sec:order_parameter}.
In the two CEF levels, there are magnetic dipole degrees of freedom $(M_x, M_y, M_z)$ within a Kramers doublet, while higher-rank multipole degrees of freedom, such as electric quadrupoles $(Q_{u(=3z^2-r^2)}, Q_{\varv(=x^2-y^2)}, Q_{yz}, Q_{zx}, Q_{xy})$, exist between two CEF levels as the interorbital degrees of freedom~\cite{jpsj_yatsushiro2020odd,yatsushiro2020nqr}.
By taking into account such dipole and quadrupole degrees of freedom, 
the model Hamiltonian is constructed as follows:
\begin{align}
\label{eq:model}
\mathcal{H}=& \mathcal{H}_{\rm CEF} + \mathcal{H}_{\rm Zeeman} + \mathcal{H}_{\rm int},\\
\label{eq:model_CEF}
\mathcal{H}_{\rm CEF} =& \Delta \sum_{r}  \sum_{\sigma=\uparrow \downarrow}   f_{r\Gamma_6\sigma}^\dagger f_{r\Gamma_6\sigma}, \\
\label{eq:model_Zeeman}
 \mathcal{H}_{\rm Zeeman} =& -g\mu_{\rm B}\sum_{r}
 {\bm H} \cdot \hat{\bm J}_{r},\\
 \label{eq:model_interaction}
\mathcal{H}_{\rm int} =& 
D \sum_{\braket{rr'}}^{\rm n.n.} \left[
\delta_u \hat{Q}_{u,r}\hat{Q}_{u,r'} 
+ \delta_v \hat{Q}_{\varv,r}\hat{Q}_{\varv,r'} 
+ \delta_{xy} \hat{Q}_{xy,r}\hat{Q}_{xy,r'}
 \right. \notag\\
&\left.
+\delta_{\rm E^+} \left(\hat{Q}_{yz,r}\hat{Q}_{yz,r'} +\hat{Q}_{zx,r}\hat{Q}_{zx,r'} \right)  
 \right. \notag\\
&\left. + \delta_{\rm E^-} \left(\hat{M}_{x,r} \hat{M}_{x,r'} + \hat{M}_{y,r} \hat{M}_{y,r'} \right) 
+ \delta_{z}\hat{M}_{z,r} \hat{M}_{z,r'} 
\right],
\end{align}
where $f_{r \Gamma_6\sigma}^\dagger$ ($f_{r \Gamma_6\sigma}$) is the creation (annihilation) operator of the $f$ electron with the quasi spin $\sigma=\uparrow, \downarrow$ in the $\Gamma_6$ level at the $r$th atomic site. 
The first term in Eq.~\eqref{eq:model} is the CEF level splitting, where we set $\Delta=90$~K from the CEF Hamiltonian in Appendix~\ref{sec:CEF}. 
The second term in Eq.~\eqref{eq:model} is the Zeeman term, where $g(=6/7)$ is the $g$ factor and $\mu_{\rm B}$ is the Bohr magneton set as $\mu_{\rm B} \to \mu_{\rm B}/k_{\rm B} = 0.67$~[K/T].
The Boltzmann factor $k_{\rm B}$ is set as $1$ in the following.
$\hat{\bm J}_{r}$ is the total angular momentum at the $r$th atomic site, where the matrix element for the second-excited CEF level is ruled out.

The last term in Eq.~\eqref{eq:model} is the antiferroic multipole-multipole interaction $(D>0)$ in the channels of electric quadrupole and magnetic dipole, where the summation is taken for the upper- and lower-nearest-neighboring Ce (A) and Ce (B) sites, $\braket{rr'}$ [see also Fig.~\ref{fig:fig1}(a)].
$\delta_{X}$ ($X=u, \varv, xy, {\rm E}^+, {\rm E}^-, z$) stands for the weight of the multipole-multipole interaction ($0 \leq \delta_{X} \leq 1$).
The electric quadrupole and magnetic dipole degrees of freedom at $r$th atomic site, $\hat{Q}_{\mu,r}$ ($\mu=u,\varv,yz,zx,xy$) and $\hat{M}_{\nu,r}$ ($\nu=x,y,z$), are given by
\begin{align}
\label{eq:E2_1}
\hat{Q}_{u,r} &= \sum_{\sigma l} c^{l}_u f_{r l\sigma}^\dagger f_{rl\sigma}, \\
\label{eq:E2_2}
\hat{Q}_{\varv,r} &= \frac{1}{2} \sum_{\sigma ll'} \tau_x^{ll'} f_{rl\sigma}^\dagger  f_{rl'\sigma}, \\
\label{eq:E2_3}
\hat{Q}_{xy,r} &= \frac{1}{2} \sum_{\sigma\sigma' ll'}  \tau_x^{ll'} \sigma_z^{\sigma\sigma'} f_{rl\sigma}^\dagger f_{rl'\sigma}, \\
\label{eq:E2_4}
\hat{Q}_{yz(zx),r} &= \frac{1}{2} \sum_{\sigma\sigma' ll'}  \tau_y^{ll'} \sigma_{x(y)}^{\sigma\sigma'} f_{rl\sigma}^\dagger f_{rl'\sigma}, \\
\label{eq:M1_1}
\hat{M}_{x(y),r} &= \sum_{\sigma\sigma'll'} \left[ c^{l}_{x(y)}  \sigma_{x(y)}^{\sigma\sigma'}  \delta_{ll'} + c'_{x(y)} \tau_x^{ll'} \sigma_{x(y)}^{\sigma\sigma'} 
 \right] f_{rl\sigma}^\dagger f_{rl'\sigma'} ,\\
\label{eq:M1_2}
\hat{M}_{z,r} &= \sum_{\sigma\sigma'l} c^{l}_{z} \sigma_{z}^{\sigma\sigma'} f_{rl\sigma}^\dagger f_{rl\sigma'},
\end{align}
where $\sigma_\mu$ and $\tau_\mu$ ($\mu=x,y,z$) are the Pauli matrices for the quasi spin $\uparrow, \downarrow$ and orbital $\Gamma_7, \Gamma_6$, respectively.
 $c_{\mu}^l$ ($\mu=u,x,y,z; l=\Gamma_6, \Gamma_7$) and $c_{x(y)}'$ are the linear combination coefficients, which are determined by the CEF parameters as $(c_u^{\Gamma_7}, c_u^{\Gamma_6})=(0.24,-0.25)$, $(c_{x,y}^{\Gamma_7}, c_{x,y}^{\Gamma_6})=(-0.14,0.24)$, $(c_{z}^{\Gamma_7}, c_{z}^{\Gamma_6})=(0.55,0.41)$, and $c_{x(y)}^{\prime}=\mp 0.28$ [upper(lower) sign is taken for $x$($y$)]. 
  The multipoles in Eqs.~\eqref{eq:E2_1}--\eqref{eq:M1_2} are normalized to be ${\rm Tr}[XX^\dagger]=1$.
By applying the Hartree approximation for $\mathcal{H}_{\rm int}$ in Eq.~\eqref{eq:model_interaction} as
\begin{align}
D \sum_{\braket{rr'}}^{\rm n.n.}
\delta_X \hat{X}_{r}\hat{X}_{r'} 
\to 
Dz \sum_{R=1}^{N} \delta_{X}(\braket{\hat{X}_{\rm A}}\hat{X}_{R{\rm B}} + \braket{\hat{X}_{\rm B}}\hat{X}_{R{\rm A}} -\braket{\hat{X}_{\rm A}}\braket{\hat{X}_{\rm B}}),
\end{align}
the mean-field Hamiltonian is obtained.
$\hat{X}_{R{\rm A(B)}}$ stands for the multipole $X$ at the sublattice A(B) in the $R$th unit cell and 
$\braket{\hat{X}_{\rm A(B)}} \equiv \sum_n \braket{n|\hat{X}_{\rm A(B)}|n} \exp{(-\beta E_n)}/Z$ is the thermal expectation value, where $\ket{n}$ and $Z$ are the eigenstate with energy $E_n$ and a partition function, respectively.
We set $\bar{D} \equiv D z$ in the following discussion.

\section{Stability of Multipolar Ordered Phases \label{sec:stability}}

We investigate the stability of the AFM and AFQ states by using the two-sublattice self-consistent mean-field calculation for the localized model in Eq.~\eqref{eq:model}.
In Sects.~\ref{sec:stability_zero_magnetic} and \ref{sec:stability_magnetic}, the stability in zero and finite magnetic fields is discussed, respectively.

The stabilization of the ordered states in the model in Eq.~(\ref{eq:model}) largely depends on the interaction parameters: $D$ and six independent $\delta_{X}$.
In this section, we consider the in-plane AFM order as the III phase, which was proposed by the neutron diffraction measurement~\cite{nikitin2020gradual}, by setting $\delta_{z}=0.3\delta_{{\rm E}^-}$ and $\delta_\varv=0.8$.
In addition, among the AFQ ordered states for the II phase, we examine the stability of the $Q_{zx}$($Q_{yz}$)-type AFQ order by setting $\delta_{{\rm E}^+}=1$, as the fourfold-rotational symmetry breaking has been recently pointed out by the x-ray diffraction measurement~\cite{matsumura2022structural}.
In addition, we set $\delta_{xy}=0$ and choose $\delta_u$, $\delta_{\rm E^-}$, and $D$ to reproduce the transition temperatures of the AFQ and AFM phases in experiments, as detailed in the subsequent section. 
We discuss the cases of the other AFM and AFQ order parameters in Sect.~\ref{sec:order_parameter}.

\subsection{Zero magnetic field \label{sec:stability_zero_magnetic}}
We start by discussing the transition temperature of the $Q_{zx}$($Q_{yz}$)-type AFQ order in a zero magnetic field. 
Let us first consider a simplified situation by ignoring the AFM interaction, i.e., $\delta_{{\rm E}^-}=\delta_z=0$.
We show the phase diagram while changing temperature ($T$) and $\bar{D}$ in Fig.~\ref{fig:fig1}(c) for three $\delta_u=0,1$, and $2$.
In the colored region above the bold line, the AFQ ordering is stabilized for each $\delta_u$.
For $\delta_u=0$, the large value $\bar{D}\sim 2\Delta$ is needed to stabilize the AFQ state, since the two CEF levels are separated by $\Delta$ in the paramagnetic state [left panel of Fig.~\ref{fig:fig1}(b)].
Once the AFQ ordering occurs, $\Gamma_7$ and $\Gamma_6$ levels mix as shown in the middle panel of Fig.~\ref{fig:fig1}(a) while keeping the two-fold degeneracy due to the time-reversal symmetry.
This result of $\bar{D}$ is consistent with that in a cubic system with a Kramers pair as a ground state~\cite{doi:10.1143/JPSJ.53.1809, comment_CeCoSi_cubic}; the AFQ phase transition is caused when the mean-field AFQ interaction $\bar{D}$ is comparable to the twice of the CEF level splitting $\Delta$ 
(See Appendix~\ref{sec:D_T0} in detail).
Meanwhile, nonzero $\delta_u$ suppresses the critical value of $\bar{D}$ as shown in Fig.~\ref{fig:fig1}(c), which is in contrast to the cubic case.
This is because the $\delta_u$ term renormalizes the tetragonal CEF level splitting effectively through the development of the ferroic $Q_u$ moment.
The renormalization of the CEF level splitting depends on the temperature because the ferroic $Q_u$ moment also depends on the temperature~\cite{comment_CeCoSi_FQ}.
In the following discussion, we adopt $\delta_u=2$ and $\bar{D}=80$, which gives a similar transition temperature to that of the II phase observed in CeCoSi, i.e., $T_{0} \simeq 12$~\cite{tanida2019successive}.

Note that the renormalized CEF level splitting near $T_0$ was not observed in the inelastic neutron scattering experiment for polycrystal under ambient pressure~\cite{nikitin2020gradual}.
However, as the quadrupolar ordering itself was not observed in that experiment,
it would be desired to obtain the temperature dependence of the CEF levels by performing the same experiment for the single crystal or under pressure to exhibit the nonmagnetic orderings (II phase), which will provide the information 
whether the renormalization of the CEF level splitting occurs or not near $T_0$  in CeCoSi.

Next, we take into account the AFM interaction to describe the AFQ-AFM phase transition.
Nonzero $\delta_{{\rm E}^-}$ replaces the AFQ ground state with the AFM one by lifting the remaining Kramers degeneracy as shown in the right panel of Fig.~\ref{fig:fig1}(b).
Figure~\ref{fig:fig1}(d) shows the phase diagram while changing $T$ and the ratio of the AFM and AFQ interactions, $\delta_{{\rm E}^-}$ and $\delta_{{\rm E}^+}$.
The solid (dotted) line means the second(first)-order phase transition.
For the small $\delta_{{\rm E}^-}/\delta_{{\rm E}^+} \lesssim 0.35$, a decrease of $T$ leads to the phase transition from the $Q_{zx}$($Q_{yz}$)-type AFQ ordering to the $M_x$($M_y$)-type AFM ordering while possessing the $Q_{zx}$($Q_{yz}$)-type AFQ moment as schematically shown in Fig.~\ref{fig:fig1}(d).
In $0.35 \lesssim \delta_{{\rm E}^-}/\delta_{{\rm E}^+} \lesssim 0.9$, the above AFM phase shows the further first-order phase transition to the ($M_x+M_y$)[($M_x-M_y$)]-type  AFM ordering with the ($Q_{yz}-Q_{zx}$)[($Q_{yz}+Q_{zx}$)]-type AFQ moment.
There are two differences in these two AFM phases; one is the in-plane anisotropy between $[100]$ and $[110]$ directions, and the other is the difference in angle relative to the AFQ moment.
Especially, the latter difference results in the different symmetry between the two types of AFM phases; the former is $2'm'm$ or $m'2'm$ and the latter is $2mm$ or $m2m$.
For the large $\delta_{{\rm E}^-}/\delta_{{\rm E}^+} \gtrsim 0.9$, only the AFM phase appears without the AFQ moment. 
In short, the interaction parameters to satisfy $\delta_{\rm E^-}/\delta_{\rm E^+}\lesssim 0.9$ reproduces the situation observed in CeCoSi except for the phase transition between two AFM phases; both AFQ and AFM phases appear while changing the temperature. 
We set $\delta_{\rm E^-}=0.7 \delta_{\rm E^+}$ in the following calculation.

\begin{figure*}[t!]
\centering
\includegraphics[width=180mm]{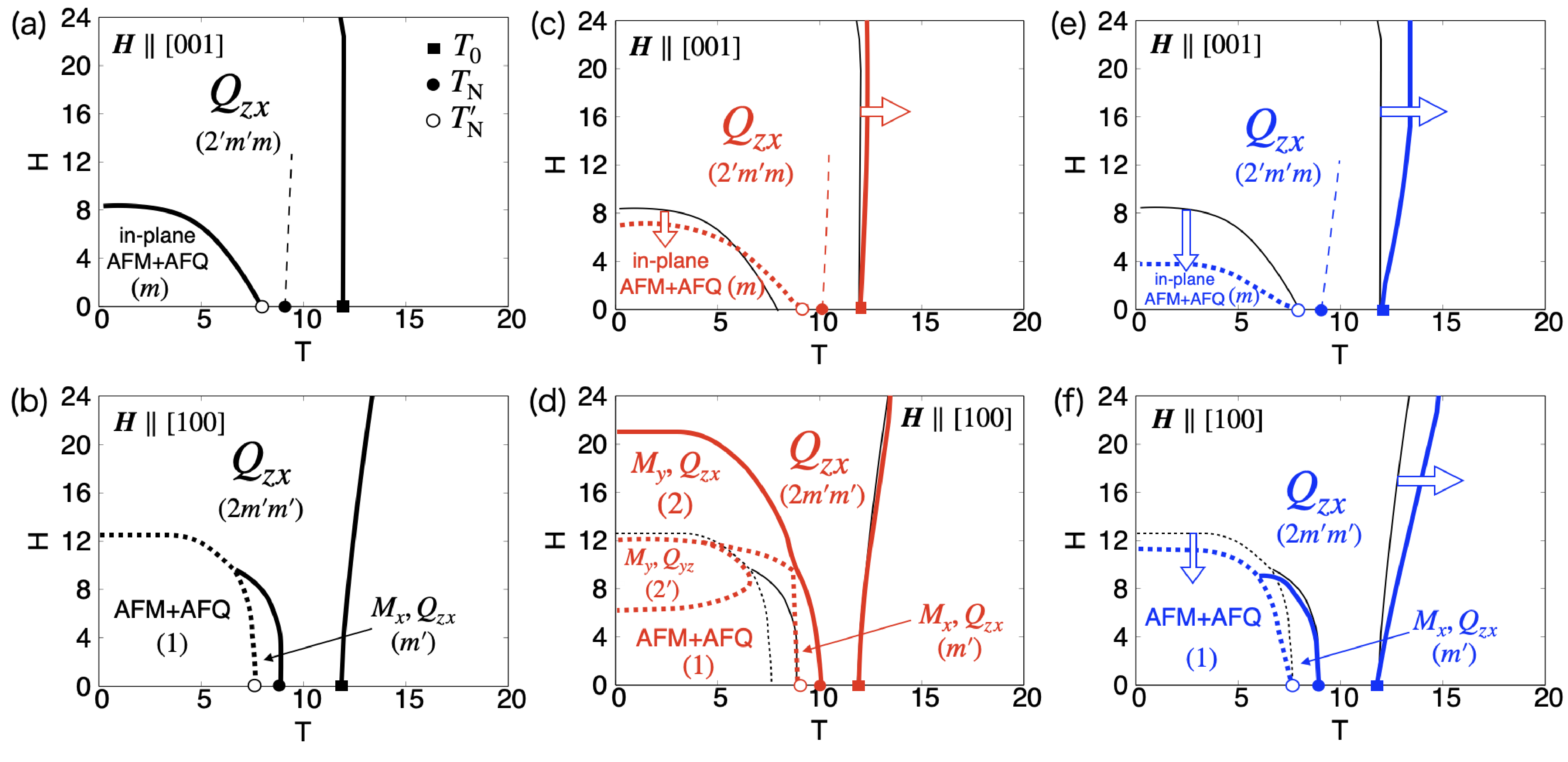}
\caption{(Color online)
(a--f) $H$-$T$ phase diagrams in the (a,c,e) [001] and (b,d,f) [100] magnetic fields.
In addition to the Zeeman coupling in (a) and (b),
the magnetic octupolar interaction $\delta_{{\rm E}_{3\beta}^-}$ is additionally considered in (c) and (d), and the effective multipolar coupling between the magnetic dipole and AFQ moment $\delta'$ is considered in (e) and (f).
The solid (dotted) line stands for the second(first)-order phase transition.
The dashed line in the [001] magnetic field represents the minimum in the $T$ derivative of the magnetization (see the main text in detail.).
The phase boundaries in (a)[(b)] is shown by the thin black lines in (c) and (e) [(d) and (f)] for reference.
 \label{fig:fig2}}
\end{figure*}

\subsection{Finite magnetic field \label{sec:stability_magnetic}}

We also investigate the AFQ and AFM phases and their phase transitions in a magnetic field.
Figures~\ref{fig:fig2}(a) and \ref{fig:fig2}(b) are the phase diagrams against the magnetic field ($H$)  and $T$, where the Zeeman field in Eq.~\eqref{eq:model} is directed along the [001] and [100] directions, respectively.
The solid (dotted) line represents the phase boundary characterized by the second(first)-order phase transition.
The filled square $(T_0)$, filled circle ($T_{\rm N}$), and empty circle ($T'_{\rm N}$) in a zero magnetic field stand for the AFQ transition temperature, AFM transition temperature, and phase transition between two types of the AFM states, respectively.

In the [001] magnetic field, the $Q_{zx}$($Q_{yz}$)-type AFQ order is stable up to the high-field region. 
The magnetic point group symmetry is $2'm'm$ ($m'2'm$) as presented in the parentheses in Fig.~\ref{fig:fig2}(a).
The AFQ transition temperature is almost the same when applying the [001] magnetic field.
The phase transition at $T_{\rm N}$ in a zero magnetic field disappears in the [001] magnetic field because of the same symmetry as the AFQ phase; a broad peak structure in the $T$ derivative of the magnetization and heat capacity is found, which is presented by the thin dashed line in Fig.~\ref{fig:fig2}(a).
On the other hand, the low-temperature AFM phase with the $(M_x+M_y)$[$(M_x-M_y)$]-type AFM moment and the $(Q_{yz}-Q_{zx})$[$(Q_{yz}+Q_{zx})$]-type AFQ moment in a zero magnetic field has the symmetry $m$ and remains in the [001] magnetic field as presented in Fig.~\ref{fig:fig2}(a).
By applying the magnetic field, these vertically coupled AFM and AFQ moments rotate continuously in the $xy$ plane and reach the $M_x$($M_y$)-type AFM and $Q_{zx}$($Q_{yz}$)-type AFQ moments coupled in parallel, which means the phase transition to the AFQ phase in the high-temperature region.

The phase diagram in Fig.~\ref{fig:fig2}(b) shows that the $Q_{zx}$-type AFQ ordering in a [100] magnetic field extends to the high-field region in the intermediate temperature region. 
It is noted that $Q_{yz}$-type AFQ state has higher energy than $Q_{zx}$-type AFQ state owing to the symmetry lowering to the orthorhombic symmetry under the [100] magnetic field.
In contrast to the result in the [001] magnetic field, the AFQ transition temperature shows a slight enhancement by the magnetic field.
This difference is attributed to the different matrix elements of $\hat{J}_{x(y)}$ and $\hat{J}_{z}$ determined by the CEF parameters; the former has the interorbital matrix element to assist the interorbital orderings like a quadrupolar order, while the latter has no interorbital component. 
In the low-temperature region under the [100] magnetic field, two types of the AFM phases survive as the different phases as shown in Fig.~\ref{fig:fig2}(b): one with the symmetry $m'$ and the other with the symmetry $1$.

The present result showing the robustness of the AFQ and AFM phases in the magnetic field is consistent with that observed in CeCoSi~\cite{tanida2019successive}.
Meanwhile, there are still several differences between them; 
one of the qualitative differences is the behavior of the AFQ transition temperature when applying the magnetic field. 
The present result shows that the AFQ transition temperature is almost unchanged for the [001] field direction, whereas the enhancement tendency was observed in experiments~\cite{tanida2019successive}.

To get an insight into the magnetic field dependence of the AFQ transition temperature, we consider two scenarios.
First, we additionally consider the antiferroic octupolar (AFO) interaction that is neglected in Eq.~(\ref{eq:model_interaction}), which is given by
\begin{align}
\label{eq:AFO}
&\mathcal{H}_{\rm AFO}^{\rm MF} \notag\\
&= \delta_{{\rm E}_{3\beta}^-} \bar{D} \sum_{R=1}^{N} \sum_{\mu=x,y}\left( \braket{\hat{M}_{\mu, {\rm A}}^\beta} \hat{M}_{\mu, {\rm B}}^\beta +   \braket{\hat{M}_{\mu,{\rm B}}^\beta} \hat{M}_{\mu, {\rm A}}^\beta -  \braket{\hat{M}_{\mu, {\rm A}}^\beta} \braket{\hat{M}_{\mu,{\rm B}}^\beta}  \right),
\end{align}
where 
\begin{align}
\label{eq:M3}
\hat{M}_{\mu,r}^\beta = \sum_{\sigma\sigma'll'} \left[ c^{l}_{\mu\beta}  \sigma_{\mu}^{\sigma\sigma'}  \delta_{ll'} + c'_{\mu\beta} \tau_x^{ll'} \sigma_{\mu}^{\sigma\sigma'}  \right] f_{rl\sigma}^\dagger f_{rl'\sigma'},
\end{align}
is the magnetic octupolar operator for $\mu=x,y$, which is different from the magnetic dipolar and electric quadrupolar operators.
The linear combination coefficients are determined by the CEF parameters as $(c_{\mu\beta}^{\Gamma_7}, c_{\mu\beta}^{\Gamma_6}, c_{\mu\beta}^\prime ) = (-0.048, -0.14, 0.454)$. 
The AFO interaction is regarded as an effective coupling between the ferroic magnetic dipole and AFQ moments, since the AFO moment can be understood as their product~\cite{shiina1997magnetic}. 
We show the modified phase diagram for $\delta_{{\rm E}^-_{3\beta}}=0.5$ in the [001] and [100] magnetic fields in Figs.~\ref{fig:fig2}(c) and \ref{fig:fig2}(d), respectively.
In the [001] magnetic field, the AFO interaction slightly increases the AFQ transition temperature as shown in Fig.~\ref{fig:fig2}(c), while it suppresses the critical field of the AFM phase.
This is because $\delta_{{\rm E}^-_{3\beta}}$ affects the stability of the in-plane AFM phase due to the same symmetry of $M_\mu^\beta$ and $M_\mu$ for $\mu=x,y$.

In the [100] magnetic field, $\delta_{{\rm E}^-_{3\beta}}$ hardly affects the AFQ transition temperature, since $M_\mu^\beta$ ($\mu=x,y$) has a different symmetry from the coupling between the ferroic $M_x$ moment and the antiferroic $Q_{zx}$ moment.
Meanwhile, the octupolar interaction changes the AFM phase drastically, 
which leads to additional three phases with different combinations of the AFM and AFQ moments.
The symmetry in each AFM phase is presented in Fig.~\ref{fig:fig2}(d).
In summary, the AFO interaction increases the AFQ transition temperature in a magnetic field but is insufficient to reproduce the phase boundaries in CeCoSi.

The second scenario is the introduction of an effective direct coupling between the uniform magnetic dipole and AFQ moments under the magnetic field.
The additional effective coupling in the [001] magnetic field with the $Q_{zx}$-type AFQ moment is given within a mean-field level by
\begin{align}
\mathcal{H}^{\rm eff}&=\delta' \bar{D} \sum_{R=1}^{N}
\left[
M_z^{\rm FM} (\hat{Q}_{zx,R{\rm A}}-\hat{Q}_{zx,R{\rm B}}) 
\right.\notag\\
& \left. 
\qquad\qquad+ Q_{zx}^{\rm AFQ}(\hat{M}_{zR,{\rm A}}+\hat{M}_{zR,{\rm B}})
-M_z^{\rm FM} Q_{zx}^{\rm AFQ}
\right],
\label{eq:H_eff}
\end{align}
where $M_z^{\rm FM} \equiv \braket{\hat{M}_{z,{\rm A}}}+\braket{\hat{M}_{z,{\rm B}}}$ and $Q_{zx}^{\rm AFQ} \equiv \braket{\hat{Q}_{zx,{\rm A}}}-\braket{\hat{Q}_{zx,{\rm B}}}$ are the ferromagnetic (FM) and the AFQ moments, respectively.

Figure~\ref{fig:fig2}(e) represents the phase diagram in the [001] magnetic field for $\delta' = -0.005$.
The result shows that the direct coupling in Eq.~\eqref{eq:H_eff} leads to the strong enhancement of the AFQ transition temperature by the magnetic field.
Meanwhile, the critical field of the AFM phase tends to be suppressed as well as that in the case of the AFO interaction in Eq.~(\ref{eq:AFO}).
Such suppression of the critical field is evaded by taking into account an additional interaction, e.g., ferroic interaction for the $xyz$-type magnetic octupole between the nearest-neighbor sites in the present AFM state, although we omit the result here.

In the [100] magnetic field, we consider a different type of the effective coupling where the magnetic dipole moment along the $z$ direction in Eq.~\eqref{eq:H_eff} is replaced to one along the $x$ direction, i.e., $M_z^{\rm FM} \to M_x^{\rm FM}$.
The phase diagram for $\delta' = -0.005$ is shown in Fig.~\ref{fig:fig2}(f).
The result shows that the phase boundary between the AFQ and paramagnetic phases is modulated to the high-temperature side with an increase of the magnetic field, whereas the AFM critical field is slightly suppressed in the presence of the effective coupling.
Thus, the effective coupling induced under the magnetic field is one of the important factors to reproduce the $H$ dependence of the AFQ transition temperature observed in CeCoSi.

\section{Susceptibility \label{sec:susceptibility}}

We investigate the behaviors of the magnetic and quadrupolar susceptibilities in the multipolar orderings while changing the temperature.
We calculate them by using the following isothermal susceptibility
\begin{align}
\chi_X (T)
&=2\sum_{nm}w_n
\frac{\left|\braket{n|\hat{X}|m}\right|^2}{E_m-E_n} \notag\\
&\quad+\frac{1}{k_{\rm B}T}
\left[
\sum_nw_n\braket{n|\hat{X}|n}^2-\left(\sum_n w_n
\braket{n|\hat{X}|n}\right)^2\right],
\end{align}
where $\ket{n}$ is the electronic state with the eigen energy $E_n$ and $w_n=e^{-\frac{E_n}{k_{\rm B}T}}$ is the Boltzmann weight of eigenstate $n$.
For magnetic and quadrupolar susceptibilities, $\chi^{\rm D}_\mu$ ($\mu=x,y,z$) and $\chi^{\rm Q}_\nu$ ($\nu = u,\varv,yz,zx,xy$), we set $\hat{X}=g \mu_{\rm B} \hat{J}_{\mu,i}$ ($i={\rm A}, {\rm B}$) and $\hat{Q}_{\nu,i}$, respectively.
In the following, we show the susceptibilities in the total two-sublattice system.
We discuss the behavior of the magnetic susceptibility in Sect.~\ref{sec:susceptibility_magnetic}.
Then, we show the temperature and field dependences of the quadrupolar susceptibility in Sect.~\ref{sec:susceptibility_quadrupole}.

\begin{figure}[htb!]
\centering
\includegraphics[width=85mm]{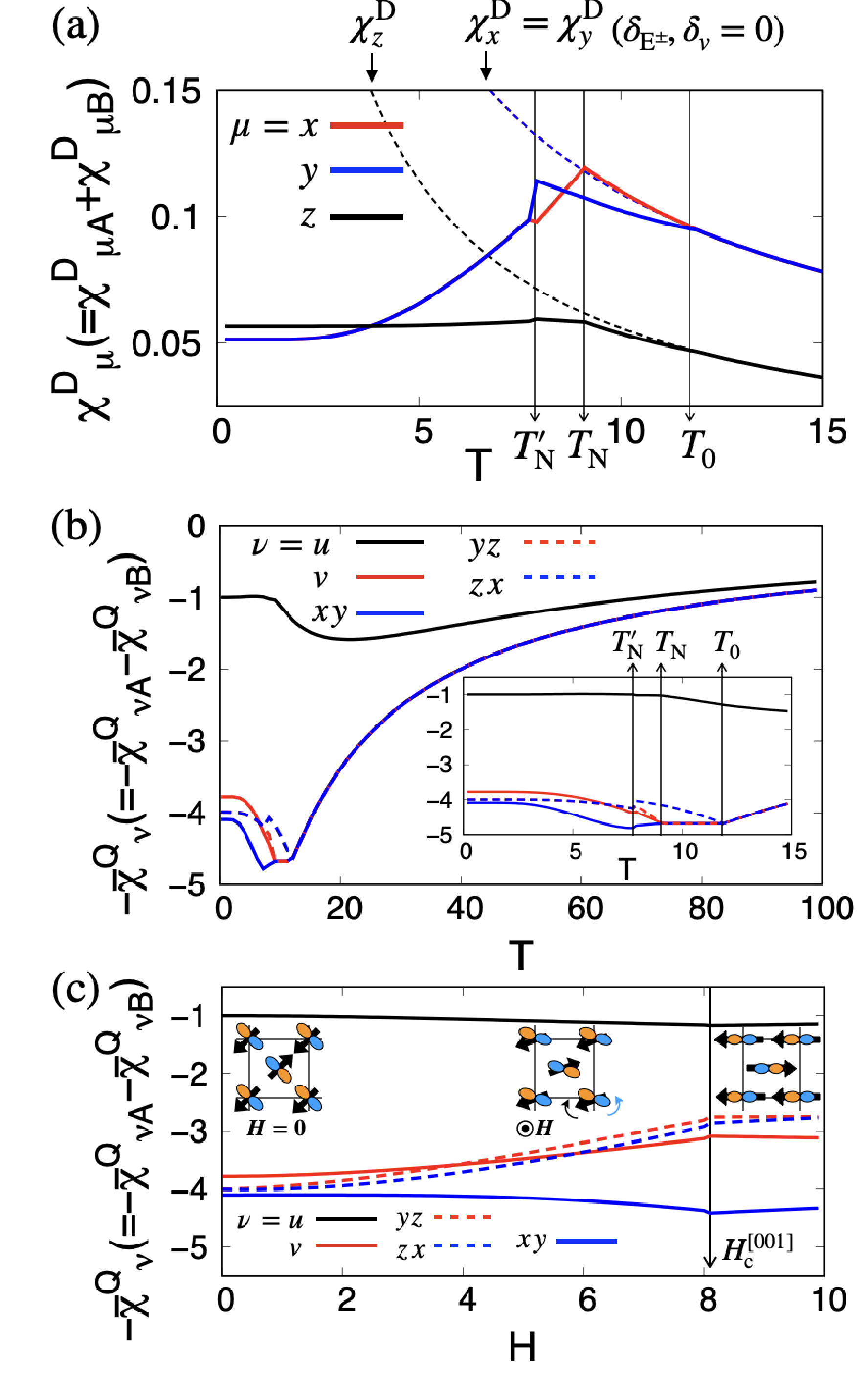}
\caption{(Color online)
(a, b) $T$ dependence of the (a) magnetic and (b) quadrupolar susceptibilities at a zero magnetic field.
(c) $H$ dependence of the quadrupolar susceptibility in the [001] magnetic field at $T=2$.
\label{fig:fig3}}
\end{figure}

\subsection{Magnetic susceptibility \label{sec:susceptibility_magnetic}}

First, we discuss the magnetic susceptibility $\chi^{\rm D}_\mu$ ($\mu=x, y, z$) in a zero magnetic field in Fig.~\ref{fig:fig3}(a).
The interaction parameters are the same as those in Fig.~\ref{fig:fig2}(a).
The magnetic susceptibilities without any electronic ordered phases for $\delta_{\rm E^\pm}=\delta_\varv=0$ are also shown by the broken lines for reference.

As shown in Fig.~\ref{fig:fig3}(a), $\chi^{\rm D}_\mu$ shows a slight anomaly at the $Q_{zx}$-type AFQ transition temperature $T_0$; $\chi^{\rm D}_x$ shows a little upturn modulation from that in the paramagnetic phase, whereas $\chi^{\rm D}_y$ and $\chi^{\rm D}_z$ show the down-turn modulation below $T_0$.
At $T_{\rm N}$, $\chi^{\rm D}_x$ shows a cusp-like anomaly as the conventional AFM order, while $\chi^{\rm D}_y$ shows almost no anomaly.
Below $T_{\rm N}'$, the behavior of $\chi_x^{\rm D}=\chi_y^{\rm D}$ is found due to the transition in terms of the AFM moment direction from the $[100]$ direction to the $[110]$ direction.
Meanwhile, $\chi^{\rm D}_z$ has an almost constant value below $T_{\rm N}$.

We focus on the behavior of $\chi^{\rm D}_\mu$ just below $T_0$.
The up or down-turn behavior of $\chi^{\rm D}_\mu$ depends on the magnitude of the effective magnetic dipole within the ground-state Kramers doublet.
To demonstrate that, we calculate a quantity of $\delta M_\mu \equiv {\rm Tr}[M_\mu^2]_{\rm AFQ} - {\rm Tr}[M_\mu^2]_{\rm para}$, where ${\rm Tr}[M_\mu^2]_{\rm AFQ}$ is calculated for the Kramers doublet with nonzero but small $Q_{zx}$ moment and ${\rm Tr}[M_\mu^2]_{\rm para}$ is for the CEF ground state.
It is approximately given by
\begin{align}
\delta M_x&= \frac{2(c'_x)^2 - 4c^{\Gamma_7}_x (c^{\Gamma_7}_x+c^{\Gamma_6}_x)}{\Delta^2}q_{zx}^2 + O(q_{zx}^4), \\
\delta M_y &= -\frac{4  c^{\Gamma_7}_y(c^{\Gamma_7}_y-c^{\Gamma_6}_y)}{\Delta^2}q_{zx}^2 + O(q_{zx}^4), \\
\delta M_z &=-\frac{4  c^{\Gamma_7}_z (c^{\Gamma_7}_z+c^{\Gamma_6}_z) }{\Delta^2}
q_{zx}^2+O(q_{zx}^4),
\end{align}
where $q_{zx} = \bar{D}\delta_{\rm E^+}\braket{\hat{Q}_{zx}}$.
By substituting the present parameters into the above expressions, one finds $\delta M_x>0$ and $\delta M_{y,z}<0$ within the second order of $q_{zx}$. 
This sign difference results in the up-turn behavior in $\chi^{\rm D}_x$ and the down-turn behavior in $\chi^{\rm D}_{y,z}$, as shown in Fig.~\ref{fig:fig3}(a).
Thus, the behavior of the magnetic susceptibility in the AFQ orderings gives information about the AFQ order parameters.
For example, one finds that the up-turn behavior of $\chi^{\rm D}_{x(y)}$ and down-turn behavior of  $\chi^{\rm D}_{y(x)}$ suggests the nonzero $Q_{zx}$ ($Q_{yz}$) moment by calculating $\delta M_\mu$ for each quadrupole moment in the present CEF parameter.

\subsection{Quadrupolar susceptibility \label{sec:susceptibility_quadrupole}}
We discuss the quadrupolar susceptibility $\chi^{\rm Q}_\nu$ ($\nu = u,\varv,yz,zx,xy$).
Figure~\ref{fig:fig3}(b) shows the quadrupolar suceptibility $-\bar{\chi}^{\rm Q}_\nu$ scaled as $\bar{\chi}^{\rm Q}_u=1$ at $T=0.2$, in a zero magnetic field.
The low-temperature region is presented in the inset of Fig.~\ref{fig:fig3}(b).
All $\chi^{\rm Q}_\nu$ components show the softening with decreasing $T$ in the paramagnetic phase.
While decreasing $T$, $\chi^{\rm Q}_\nu$ shows different modulations in the AFQ ordered phase depending on the quadrupole component, as found
in the conventional AFQ ordered systems like CeB$_{6}$~\cite{nakamura1994quadrupole}; 
$\chi^{\rm Q}_{zx}$ shows a cusp-like anomaly at $T_0$ and upturns with decreasing $T$, while $\chi^{\rm Q}_{\varv}$, $\chi^{\rm Q}_{yz}$, and $\chi^{\rm Q}_{xy}$ show almost constant values. 
In the AFM phases, three components of $-\bar{\chi}^{\rm Q}_{\nu'}$ ($\nu'=\varv,yz,zx$) mostly show the up-turn behavior with decreasing $T$, although $-\bar{\chi}^{\rm Q}_{xy}$ slightly decreases between $T_{\rm N}$ and $T_{\rm N}'$.
On the other hand, $\chi^{\rm Q}_{u}$ shows a broad peak around $T\sim15$, which roughly corresponds to half of the effective CEF splitting renormalized by $\delta_u$ term in Eq.~\eqref{eq:model} and described by 
$\Delta^{\rm eff} \sim \Delta - \bar{D}\delta_u \braket{\hat{Q}_{u,{\rm A(B)}}} \sim 30$~K.
While decreasing $T$,
$\chi_u^{\rm Q}$ shows the anomaly at the AFM transition temperature and reaches the constant value.

Moreover, we investigate the behavior of $\chi^{\rm Q}_\nu$ in a magnetic field by focusing on the region below $T_{\rm N}$. 
We here do not consider the effects of the octupolar interaction and the effective FM-AFQ couplings introduced in Sect.~\ref{sec:stability_magnetic}, 
as they do not give a qualitative difference.
Figure~\ref{fig:fig3}(c) shows the $\chi_{\nu}^{\rm Q}$ in the $[001]$ magnetic field at $T=2$, where $H_{\rm c}^{[001]}$ is the critical field of the AFM phase.
$\chi_{yz}^{\rm Q}$ and $\chi_{zx}^{\rm Q}$ split by the magnetic field with the hardening, $\chi_{\varv}^{\rm Q}$ also shows the hardening, and  $\chi_{xy}^{\rm Q}$ and $\chi_{u}^{\rm Q}$ show the softening.
Such various behaviors are due to the rotation of the AFM and AFQ moments when increasing the magnetic field as schematically shown in Fig.~\ref{fig:fig3}(c).
When the $(Q_{yz}-Q_{zx})$[$(Q_{yz}+Q_{zx})$]-type AFQ moment with the relation $\chi_{yz}^{\rm Q}=\chi_{zx}^{\rm Q}$ rotates to the $Q_{zx}$($Q_{yz}$)-type one, $\chi_{yz}^{\rm Q}$ and $\chi_{zx}^{\rm Q}$ split and $\chi_{xy}^{\rm Q}$ ($\chi_{ \varv}^{\rm Q}$) shows softening (hardening). 
Since the rotation of the quadrupole moment by a magnetic field does not occur when the AFM and AFQ moments are parallelly coupled, the quadrupolar susceptibility in a magnetic field provides information about the way of coupling between the AFM and AFQ moments.

\section{Other Order Parameters \label{sec:order_parameter}}

Finally, we briefly discuss the result for other types of AFQ and AFM phases with ${\bm q}={\bm 0}$, which are stabilized by setting the different model parameters in Eq.~\eqref{eq:model} in the mean-field calculation.
We consider the five types of AFQ orderings activated in $\Gamma_7$-$\Gamma_6$ levels.
We also show the result for the different CEF level schemes consisting of two $\Gamma_7$ levels, where three types of AFQ and one type of antiferrohexadecapolar (AFH) orderings are possible. 
In all the cases, we list the correspondence between the order parameters and crystallographic (magnetic) point group in Table~\ref{table:table1}. 
We also present the important components in the AFO interaction or antiferroic triacontadipolar (AFT) interaction to realize the increase of $T_0$ under a magnetic field observed in the experiment. 
The checkmark $\checkmark$ means that the increase of $T_0$ occurs only by the Zeeman coupling.
Since several AFO interactions affect not only the stability of the AFQ phase but also that of the AFM phase, one might need to consider other types of coupling like that introduced in Sect.~\ref{sec:stability_magnetic} to realize the sufficient increase of $T_0$ by a magnetic field with keeping the relation $T_0>T_{\rm N}$.

Moreover, we show the symmetry of the AFM orderings with the AFQ moment in each case. 
We here consider the situation where the AFM moment lies in the $xy$ or $z$ direction.
In a zero magnetic field, we show information about uniform magnetization; any spontaneous magnetization has not been detected in a zero magnetic field experimentally~\cite{tanida2019successive}. 
Under [001] and [100] magnetic fields, the tendency of the critical field is shown compared to that of Figs.~\ref{fig:fig2}(a) and \ref{fig:fig2}(b), respectively, although the stability was not investigated exhaustively for all combinations of the parameters.
It is noted that several types of phases are unstable or have no clear phase boundary between the AFQ phase and AFQ+AFM phase because of the same symmetry, which are presented as ``---'' and ``(No phase boundary)'' in Table~\ref{table:table1}, respectively.

\tabcolsep = 3pt
\begin{table*}[h!]
\centering
\caption{
The results of various order parameters with AFQ(+AFM) moments and ordering vector ${\bm q}={\bm 0}$ are summarized.
The crystallographic point group (CPG) or magnetic point group (MPG) symmetry in each order is shown. 
In the column ``increase of $T_0$" in the AFQ order (II phase), $\checkmark$ means that an increase of $T_0$ under the magnetic field is found only by the Zeeman coupling, while AFO int. means the necessity of the additional effect, such as the AFO interaction in Eq.~(\ref{eq:AFO}) or the FM-AFQ interaction in Eq.~(\ref{eq:H_eff}).
The relevant component of the magnetic octupole is also shown.
In AFQ+AFM order (III phase), ``mag." represents the presence ($\neq 0$) or absence ($=0$) of the uniform magnetization in a zero magnetic field ($\bm{H}=\bm{0}$).
For finite magnetic fields, ``high/low/similar'' means that the critical field of the III phase tends to be higher/lower/similar when comparing that in Figs.~\ref{fig:fig2}(a) and \ref{fig:fig2}(b).
$M_{\parallel}$ represents the arbitrary AFM moment in the $xy$ plane.
``---'' represents the situation where the corresponding state is not stabilized in the present model.
``(No phase boundary)'' stands for no clear phase boundary between the AFQ phase and AFQ+AFM phase 
because of the same symmetry.
\label{table:table1}
}
\begin{tabular}{clccccclcccccc}
\hline
& \multicolumn{6}{l}{AFQ order} & \multicolumn{7}{l}{AFQ+AFM order}  \\
& type & \multicolumn{1}{c}{${\bm H}={\bm 0}$} & \multicolumn{2}{c}{${\bm H}_{\parallel [001]}$} & \multicolumn{2}{c}{${\bm H}_{\parallel [100]}$}
&\multicolumn{1}{c}{type} 
& \multicolumn{2}{c}{${\bm H}={\bm 0}$}& \multicolumn{2}{c}{${\bm H}_{\parallel [001]}$} & \multicolumn{2}{c}{${\bm H}_{\parallel [100]}$}
\\ \cline{3-3}\cline{4-5}\cline{6-7} 
\cline{9-10}\cline{11-12}\cline{13-14}

&  & CPG & MPG & increase of $T_0$ & MPG & increase of $T_0$ & & MPG & mag. & MPG & critical field & MPG & critical field \\
 \hline
$\Gamma_7$-$\Gamma_6$
& $Q_u$ & $4mm$ & $4m'm'$ & $M_z^\alpha$ AFO int.$^*$$^{\rm a}$ & $mm'2'$ & $M_x^{\alpha, \beta}$ AFO int.$^*$ & 
 $Q_u+M_z$ & \multicolumn{2}{c}{---} & \multicolumn{2}{c}{(No phase boundary)} & \multicolumn{2}{c}{---}\\
 
& & & & & & & 
 $Q_u+M_{\parallel}$ &  \multicolumn{2}{c}{---} &  \multicolumn{2}{c}{---} & $2'$ & high\\
 
\cline{2-14}
   
 & $Q_\varv$& $\bar{4}m2$ & $\bar{4}m'2'$ &$M_{z}^\beta$ AFO int. & $mm'2'$ & $M_x^{\alpha, \beta}$ AFO int.$^*$ & 
 $Q_\varv+M_z$  & $\bar{4}'m'2$ & $0$ & $m'm'2$ & high & $m'$ & low \\
 
 & & & & & & & 
 $Q_\varv+M_{\parallel}$ & \multicolumn{2}{c}{---} &  \multicolumn{2}{c}{---} & $2'$ & high \\

  \cline{2-14}
   
&  $Q_{xy}$ & $\bar{4}2m$ & $\bar{4}2'm'$ & $M_{xyz}$ AFO int. & $22'2'$ & \checkmark & 
$Q_{xy}+M_z$  & $\bar{4}'2m'$ & $0$ & $m'm'2$ & high & $2$ & high \\
 
 & & & & & & & 
 $Q_{xy}+M_{\parallel}$ & $mm'2'$$^{\rm b}$ & $\neq 0$ & \multicolumn{2}{c}{---} & $2'$ & low \\

  \cline{2-14}
   
& $Q_{zx}$ & $2mm$ &$2'm'm$ & $M_x^{\alpha, \beta}$ AFO int.$^*$ & $2m'm'$ & \checkmark & 
$Q_{zx}+M_{z}$ &
\multicolumn{2}{c}{ \multirow{2}{*}{---}}  & \multicolumn{2}{c}{ \multirow{2}{*}{---}} & \multicolumn{2}{c}{(No phase boundary)}  \\

& $Q_{yz}$ & $m2m$ &$m'2'm$ & $M_x^{\alpha, \beta}$ AFO int.$^*$ & \multicolumn{2}{c}{---} &
 $Q_{yz}+M_{z}$ & & & & &  \multicolumn{2}{c}{---} \\
 \cline{8-14}

 & & & & & & & 
$Q_{zx}+M_{y}$ &$2mm$ & $0$ & \multirow{4}{*}{$m$} & \multirow{4}{*}{Fig.~\ref{fig:fig2}(a)} & $2$  & high  \\

 & & & & & & & 
$Q_{yz}+M_{x}$ & $m2m$  & $0$ & & &\multicolumn{2}{c}{---} \\

 & & & & & & & 
$Q_{[110]}+M_{[\bar{1}10]}$ & $2mm$ & $0$ & & & \multirow{2}{*}{$1$} & \multirow{2}{*}{Fig.~\ref{fig:fig2}(b)} \\

 & & & & & & & 
$Q_{[\bar{1}10]}+M_{[110]}$ & $m2m$ & $0$ & \\
  \cline{8-14}
  
 & & & & & & & 
 $Q_{zx}+M_{x}$ & $2'm'm$ & $\neq 0$ & \multicolumn{2}{c}{\multirow{4}{*}{(No phase boundary)}} & $m'$ & low
\\
   
& & & & & & & 
 $Q_{yz}+M_{y}$ & $m'2'm$ & $\neq 0$ & & & $2'$ & similar
\\
 
& & & & & & & 
 $Q_{[110]}+M_{[110]}$ & $2'm'm$ & $\neq 0$ & & & \multirow{2}{*}{$1$} & \multirow{2}{*}{Fig.~\ref{fig:fig2}(b)}
 \\
 
 & & & & & & & 
 $Q_{[\bar{1}10]}+M_{[\bar{1}10]}$ & $m'2'm$ & $\neq 0$ 
 \\

 \hline
 $\Gamma_7$-$\Gamma_7$
& $Q_u$ & $4mm$ & $4m'm'$ & $M_z^\alpha$ AFO int.$^*$ & $mm'2'$ & $M_x^{\alpha, \beta}$ AFO int.$^*$ & 
 $Q_u+M_z$ & $4m'm'$ & $\neq 0$ &\multicolumn{2}{c}{(No phase boundary)} & $m'$ & high\\
 
& & & & & & & 
 $Q_u+M_{\parallel}$ & \multicolumn{2}{c}{---} & \multicolumn{2}{c}{---} &\multicolumn{2}{c}{(No phase boundary)}  \\

\cline{2-14}
 
& $Q_{zx}$ & $2mm$ &$2'm'm$ & \checkmark & $2m'm'$ & $M_z^\alpha$ AFO int.$^*$ & 
 $Q_{zx}+M_z$ & $2m'm'$ & $\neq 0$ & \multirow{2}{*}{$m'$} & \multirow{2}{*}{high} & \multicolumn{2}{c}{(No phase boundary)}
\\

 & $Q_{yz}$ & $m2m$ &$m'2'm$ & \checkmark & \multicolumn{2}{c}{---} & 
 $Q_{yz}+M_z$ & $m'2m'$ & $\neq 0$ & & & \multicolumn{2}{c}{---} \\

 \cline{8-14}
  & & & & & & & 
$Q_{zx}+M_{y}$ &\multicolumn{2}{c}{\multirow{4}{*}{---}} & \multicolumn{2}{c}{\multirow{4}{*}{---}}&  \multicolumn{2}{c}{\multirow{4}{*}{---}}  \\

 & & & & & & & 
$Q_{yz}+M_{x}$ & & & & & \\

 & & & & & & & 
$Q_{[110]}+M_{[\bar{1}10]}$ & & & & & \\

 & & & & & & & 
$Q_{[\bar{1}10]}+M_{[110]}$ & && \\
  \cline{8-14}
  
 & & & & & & & 
 $Q_{zx}+M_{x}$ & $2'm'm$ & $\neq 0$ & \multicolumn{2}{c}{\multirow{4}{*}{(No phase boundary)}} & $m'$& high
\\
   
& & & & & & & 
 $Q_{yz}+M_{y}$ & $m'2'm$ & $\neq 0$ & & &$2'$ & high
\\
 
& & & & & & & 
 $Q_{[110]}+M_{[110]}$ & $2'm'm$ & $\neq 0$ & & &  \multicolumn{2}{c}{\multirow{2}{*}{---}}
 \\
 
 & & & & & & & 
 $Q_{[\bar{1}10]}+M_{[\bar{1}10]}$ & $m'2'm$ & $\neq 0$ 
 \\
 
 \cline{2-14}
 
& $Q_{4z}^\alpha$ & $422$ & $42'2'$ & $M_{5u}$ AFT int. & $22'2'$ & $M_x^{\alpha,\beta}$ AFO int.$^*$ & $Q_{4z}^\alpha+M_z$ & \multicolumn{2}{c}{---} & \multicolumn{2}{c}{---} &  \multicolumn{2}{c}{---}  \\

 & & & & & & & $Q_{4z}^\alpha+M_{\parallel}$ & $2'$ & $\neq 0$ & $2'$ & high & \multicolumn{2}{c}{(No phase boundary)}\\

\hline
\\
\multicolumn{14}{l}{a: AFO interaction with notation $*$ affects AFM orderings owing to the effective coupling between octupole and dipole moments.}\\
\multicolumn{14}{l}{b: MPG symmetry $mm'2'$ represents the case when the AFM moment is parallel to [110] direction.
}\\
\end{tabular}
\end{table*}

\section{Summary \label{sec:summary}}
We investigated the stability of the multipolar orderings in the tetragonal $f$ electron material CeCoSi.
By using the mean-field calculation for the effective localized model, we clarified the finite-temperature phase transition between the AFM and AFQ ordered states in the presence of the large CEF level splitting from the ground-state Kramers doublet.
We have shown that the ($3z^2-r^2$)-type AFQ interaction plays a role in renormalizing the CEF level splitting and assists an interorbital multipolar ordering in the tetragonal symmetry.
We also examined the behavior of the AFQ and AFM phases in a magnetic field with particular attention to the field dependence of the transition temperature of the AFQ phase.
Moreover, we discussed the magnetic and quadrupolar susceptibilities in the AFQ and AFM states, which provides information about the order parameter of the quadrupolar phase and the coupling between AFQ and AFM moments.
Finally, we provided a useful table in terms of the possible AFQ and AFM states within the present model, which includes information about the symmetry of each phase, the behavior of the transition temperature, the presence/absence of magnetization, and the tendency of the stability and critical field in the mean-field calculation.
The present study will not only help the identification of the order parameter of CeCoSi but also stimulate further study of potential materials to show interorbital multipolar orderings in low-symmetry tetragonal systems.

One of the remaining issues is to clarify the microscopic origin of the multipolar interactions, which are introduced as phenomenological parameters in the present model. 
For that purpose, it is desired to construct a low-energy effective tight-binding model based on {\it ab-initio} calculations. 
In addition, it is intriguing to evaluate the effect of the local parity mixing between orbitals with the different parity in the absence of the inversion center at the atomic site~\cite{Yanase_JPSJ.83.014703,hayami2014toroidal,hayami2016emergent,hayami2018microscopic,yatsushiro2019atomic}. 
Since the site symmetry of the Ce site in CeCoSi is $4mm$
in the high-temperature I phase, the hybridization between the $d$ (or $s$) and $f$ orbitals is expected, which results in the anisotropic interactions in the dipolar and quadrupolar components~\cite{jpsj_yatsushiro2020odd, proc_yatsushiro2020antisymmetric}. 
Such an issue will be left for future study.

Finally, let us comment on the relation between the CEF level splitting and the transition temperature.
Although we propose a scenario where the CEF level is renormalized by one of the multipole-multipole interactions, it is not supported by the present experiment for polycrystal under ambient pressure~\cite{nikitin2020gradual}. 
To settle this point, a further systematic experiment for a single crystal or under pressure is highly desired as mentioned in Sect.~\ref{sec:stability}. 
If the renormalization of the CEF level splitting is not observed near the nonmagnetic ordered phase, one needs fine-tuning of the interaction as discussed in Appendix~\ref{sec:D_T0} to obtain the quadrupolar ordering with $T_0/\Delta \sim 0.1$, since it is natural that the transition temperature is comparable to the CEF level splitting ($T_0/\Delta \sim 1$) in a localized model within the mean-field calculations, as found in the other $f$-electron compound with the interorbital quadrupolar ordering YbRu$_2$Ge$_2$~\cite{PhysRevB.77.045105}. 
In such a case, additional effects like conduction electron might be important to account for the mechanism of the stabilization of the quadrupolar orderings in CeCoSi~\cite{jpsj_yatsushiro2020odd}.

\section*{Acknowledgments}
This research was supported by JSPS KAKENHI Grants Numbers JP19K03752, JP19H01834, JP21H01037, JP22H04468, JP22H00101, JP22H01183, and by JST PRESTO (JPMJPR20L8).
M.Y. is supported by a JSPS research fellowship and supported by JSPS KAKENHI (Grant No. JP20J12026).

\appendix
\section{Crystalline Electric Field \label{sec:CEF}}

\begin{figure}[htb!]
\centering
\includegraphics[width=87mm]{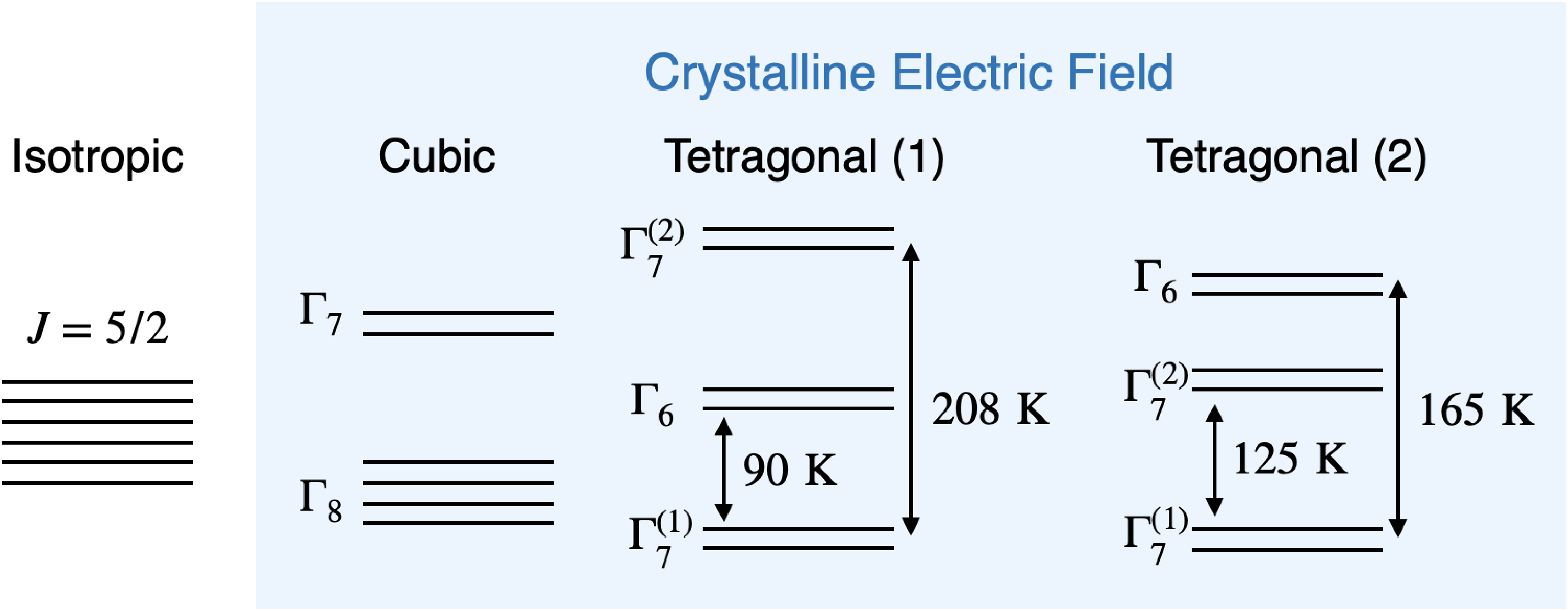}
\caption{(Color online) The level splittings of $J=5/2$ multiplet in the cubic and tetragonal crystalline electric fields (CEFs).
The CEF parameters $(B_{20}, B_{40}, B_{44})=(-0.95, -0.14, 3.8)$~K are used for ``Tetragonal (1)'', whereas $(B_{20}, B_{40}, B_{44})=(-1.26, 0.487, 1.36)$~K are used for  ``Tetragonal (2)''.
\label{fig:sm_fig1}}
\end{figure}

We show the CEF Hamiltonian and the basis wave functions of 4$f$ electron with $f^1$ configuration in the Ce$^{3+}$ ion.
The tetragonal CEF Hamiltonian at the Ce site with $4mm$ ($C_{\rm 4v}$) symmetry is given as 
\begin{align}
\label{eq:CEF_Hamiltonian}
\mathcal{H}_{\rm CEF} =& B_{20} \hat{O}_{20} + B_{40} \hat{O}_{40} + B_{44} \hat{O}_{44}^{\rm (c)},
\end{align}
where $B_{lm}$ and $\hat{O}_{lm}$ are the CEF parameter and Stevens operator~\cite{hutchings1964point}, respectively.
 $B_{60}$ and $B_{64}$ terms are omitted by supposing the $J=5/2$ basis.
We also omit the contribution of the local hybridization between $d$ and $f$ orbitals in the absence of the local inversion center at the Ce site.
In the CEF Hamiltonian in Eq.~\eqref{eq:CEF_Hamiltonian}, the sixfold $J=5/2$ basis split into one $\Gamma_6$ level and two $\Gamma_7$ levels.
The eigen energies of $\Gamma_6$ and $\Gamma_7$ levels, $E_{\Gamma_6}$ and $E_{\Gamma_7^{(1,2)}}$, are given as follows:
\begin{align}
E_{\Gamma_6}=& -8 B_{20}+120 B_{40},\\
E_{\Gamma_7^{(1)}}= & 4 B_{20}-60B_{40} -6 \sqrt{(B_{20}+20 B_{40})^2+20 B_{44}^2},\\
E_{\Gamma_7^{(2)}}= & 4 B_{20}-60B_{40} +6 \sqrt{(B_{20}+20 B_{40})^2+20 B_{44}^2},
\end{align}
whose wave functions are represented as
\begin{align}
\label{eq:CeCoSi_CEF_eigen_state_1}
\ket{\Gamma_6,\uparrow \downarrow} &= \left| \pm \frac{1}{2} \right>, \\ \
\label{eq:CeCoSi_CEF_eigen_state_2}
\ket{\Gamma_7^{(i)},\uparrow \downarrow} &= a_1^{(i)}\left| \pm \frac{5}{2} \right> + a_2^{(i)}\left| \mp \frac{3}{2} \right> \ (i=1,2),
\end{align} 
where $a_1^{(i)}$ and $a_2^{(i)}$ ($i=1,2$) are the linear combination coefficients determined by $B_{20}$, $B_{40}$, and $B_{44}$.
The matrix elements of dipole, quadrupole, and octupole in Eqs.~\eqref{eq:E2_1}--\eqref{eq:M1_2} and \eqref{eq:M3} are obtained by calculating the matrix element of each multipole for the CEF basis in Eqs.~\eqref{eq:CeCoSi_CEF_eigen_state_1} and \eqref{eq:CeCoSi_CEF_eigen_state_2} and normalizing them to be ${\rm Tr}[XX^\dagger]=1$ in the subspace of the low-energy two CEF levels.

Two types of CEF parameters are proposed by experiments: $(B_{20}, B_{40}, B_{44})=(-0.95, -0.14, 3.8)$~K~\cite{mitsumoto2019} and $(B_{20}, B_{40}, B_{44})=(-1.26, 0.487, 1.36)$~K~\cite{nikitin2020gradual}.
The former CEF parameter gives the $\Gamma_7$ ground state and the $\Gamma_6$ first excited state with 90~K level splitting denoted as ``Tetragonal (1)'' in Fig.~\ref{fig:sm_fig1}, whereas the latter leads to the $\Gamma_7$ ground and first excited states with 125~K level splitting denoted as ``Tetragonal (2)'' in Fig.~\ref{fig:sm_fig1}. 
We discuss the case of ``Tetragonal (1)'' in the main text, although that of ``Tetragonal (2)'' is also shown in Table~\ref{table:table1}. 

\section{Relation between the interaction and transition temperature in the fixed CEF levels \label{sec:D_T0}}

\begin{figure}[htb!]
\centering
\includegraphics[width=70mm]{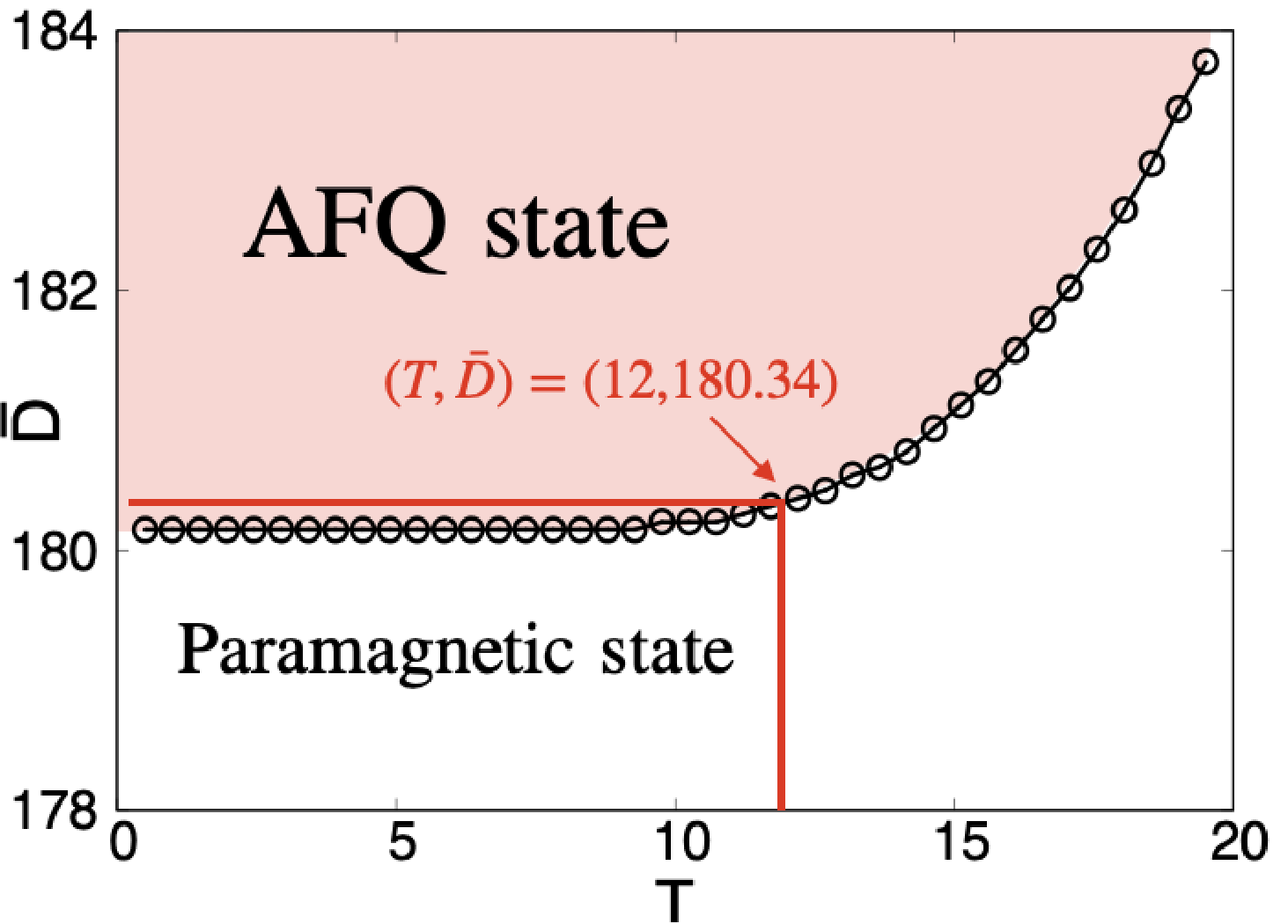}
\caption{
(Color online) The relation between multipole-multipole interaction $\bar{D}$ and AFQ transition temperature $T_0$ under the fixed CEF level splitting $\Delta=90$, where $\delta_{{\rm E}^+}=1$ and the other parameters are set to zero.
\label{fig:sm_fig2}}
\end{figure}

We show the interaction dependence of the AFQ transition temperature $T_0$ to discuss the extent of tuning of the interaction $\bar{D}$ in Eq.~\eqref{eq:model} to reproduce the transition temperature observed in experiments under the large CEF level splitting.
Figure~\ref{fig:sm_fig2} shows the critical value of $\bar{D}$ to occur the AFQ ordering in each temperature.
The model parameter is set to $\delta_{{\rm E}^+}=1$ and the CEF splitting is fixed to $\Delta=90$~K.
The other parameters are set to zero.
To reproduce $T_0=12$~K, the fine-tuning of $\bar{D}=180.34$ is needed in the absence of the temperature-dependent renormalization of the CEF level splitting.

\bibliographystyle{jpsj}
\bibliography{ref}

\begin{thebibliography}{10}
\bibitem{comment_CeCoSi_FQ}
We note that the renormalization of the tetragonal CEF splitting does not
  change for the ferroquadrupolar interaction $\delta_{\rm E^+} = -1$, which
  leads to a qualitative similar phase diagram to that in
  Fig.~\ref{fig:fig1}(c); the AFQ phase is replaced by the ferroquadrupolar
  ordered phase. Meanwhile, there is a qualitative difference in the III phase.
  When the AFM interaction is introduced, it turns into the AFM one with the
  out-of-plane component because of the monoclinic or less symmetry in the
  ferroquadrupolar ordering. Besides, the ferrooctupolar interaction is
  necessary to realize the increasing transition temperature of the quadrupolar
  phase under a magnetic field instead of the AFO interaction for the AFQ
  phase.
\end{thebibliography}

\end{document}